\documentclass[aps,twocolumn,prl,preprintnumbers]{revtex4}
\usepackage{times}
\usepackage{amsmath,amssymb}
\usepackage{enumerate}
\usepackage{color}

\newcommand{\be}{\begin{equation}}
\newcommand{\ee}{\end{equation}}
\newcommand{\ba}{\begin{align}}
\newcommand{\ea}{\end{align}}

\begin{document}

\preprint{CQUeST-2008-0203, IPMU 08-0055, OU-HET-609/2008, UTAP-603}

\title{Consistent Anti-de Sitter-Space/Conformal-Field-Theory Dual\\for a Time-Dependent Finite Temperature System}

\author{Shunichiro Kinoshita}
\email[E-mail: ]{kinoshita@utap.phys.s.u-tokyo.ac.jp}
\affiliation{Department of Physics, The University of Tokyo, Tokyo 113-0033, Japan}

\author{Shinji Mukohyama}
\email[E-mail: ]{shinji.mukohyama@ipmu.jp}
\affiliation{Institute for the Physics and Mathematics of the Universe (IPMU), The University of Tokyo, Chiba 277-8582, Japan}

\author{Shin Nakamura \footnote{Present address: Asia Pacific Center for Theoretical Physics (APCTP), POSTECH, Pohang 790-784, Korea; nakamura@apctp.org}}
\email[E-mail: ]{nakamura@hanyang.ac.kr}
\affiliation{Center for Quantum Spacetime (CQUeST), Sogang University, Seoul 121-742, Korea}

\author{Kin-ya Oda}
\email[E-mail: ]{odakin@phys.sci.osaka-u.ac.jp}
\affiliation{Department of Physics, Osaka University, Osaka 560-0043, Japan}


\begin{abstract}
We propose a consistent setup for the holographic dual of the strongly coupled large-$N_{c}$ ${\cal N}=4$ super Yang-Mills theory plasma which undergoes the Bjorken flow relevant to the quark-gluon plasma at BNL Relativistic Heavy Ion Collider and CERN LHC.
The dual geometry is constructed order by order in a well-defined late-time expansion. The transport coefficients are determined by the regularity of the geometry. We prove, for the first time, that the dual geometry has an apparent horizon hence an event horizon, which covers a singularity at the origin. Further we prove that the dual geometry is regular to all orders in the late-time expansion under an appropriate choice of the transport coefficients.
This choice is also shown to be unique. Our model serves as a concrete well-defined example of a time-dependent anti-de Sitter-space/conformal-field-theory dual.
\end{abstract}

\maketitle

The quark-gluon plasma produced at the Relativistic Heavy Ion Collider (RHIC) is an ultimate state of matter whose viscosity takes the lowest value that has ever been observed.
The plasma will also be produced with higher temperature at the Large Hadron Collider (LHC). 
An extremely small viscosity indicates that the interactions among the constituent particles are very strong.
The plasma produced by a central collision exhibits approximately a boost-invariant one-dimensional expansion, namely the Bjorken flow~\cite{Bjorken}.
It is a challenging problem to construct a theory that describes such a strongly interacting time-dependent system from the first principle.

The Anti-de Sitter space (AdS)/Conformal Field Theory (CFT) cor\-re\-spond\-ence provides a framework that accounts for a strongly coupled gauge theory in terms of a dual gravitational picture in higher dimensions~\cite{Malda,GKPW}.
Finite temperature has been introduced into the framework of AdS/CFT in Ref.~\cite{Witten:1998zw} and the deconfinement phase of the YM-theory plasma has been found to correspond to the AdS black hole.
However, the AdS/CFT cor\-re\-spond\-ence for a time-dependent system has not been fully established yet.

An attempt to construct a holographic dual of the Bjorken flow of the $\mathcal{N}=4$ Super Yang-Mills (SYM) plasma has been initiated by Janik and Peschanski~\cite{Janik-Pes} and has been further pushed forward by several groups~\cite{Nakamura-Sin,Bak-Janik,Janik-eta,HJ} (see also~\cite{Nakamura_another}).
Especially, the transport coefficients including the shear viscosity have been obtained from the regularity of the dual geometry~\cite{Janik-eta,HJ}. The dual geometry has been interpreted as a time-dependent black hole.

However, it is quite non-trivial even to show the very existence of event horizon in a time-dependent geometry. As far as the authors know, any proof of the presence of event horizon on the dual time-dependent geometry has not been reported.
Furthermore, it has been claimed in Ref.~\cite{BBHJ} that there is a logarithmic singularity at the third order of the late-time expansion in the dual geometry which cannot be removed within the framework of the ten-dimensional type IIB supergravity.

In this Letter we propose a new dual geometry which is free from the above mentioned problem. We prove the existence of event horizon, as well as the absence of naked singularity in the bulk to all orders of the late-time expansion.

{\it Bjorken flow and its holographic dual}.---Hydrodynamics tells how the stress tensor ($T_{ij}$) of the fluid evolves in time provided that the equation of state (EOS) and the transport coefficients are given. The EOS for the present case is given as the traceless condition of the stress tensor because of the conformality of the gauge theory. The Bjorken flow of conformal fluid is simple enough to solve the hydrodynamic equation. The solution is given as
\begin{align}
	T_{\tau\tau}(\tau)
		&=	\epsilon_0\left(\tau^{-4/3}-2\eta_0\tau^{-2}+\epsilon_0^{(2)}\tau^{-8/3}+\cdots\right),
			\label{T_four_D}
  \end{align}
where $\tau$ is the proper time of the fluid, $\epsilon_0$ is a free parameter which determines the initial energy density, and $\eta_0$ is proportional to the shear viscosity $\eta$: 
$\eta(\tau)=\epsilon_0\eta_0\left(T_{\tau\tau}(\tau)/\epsilon_0\right)^{3/4}$. We follow the second-order relativistic hydrodynamics of conformal fluid proposed by~Refs.~\cite{BRSSS,BHMR} in which
$\epsilon_0^{(2)}$ is given by
${9\eta_0^2+4(\lambda_1^0-\eta_0\tau_\Pi^0)\over6}$ 
where $\tau_\Pi^0$ and $\lambda_1^0$ are proportional to the relaxation time $\tau_\Pi$ and another second-order transport coefficient $\lambda_1$, respectively: 
$\tau_\Pi(\tau)=\tau_\Pi^0\left(T_{\tau\tau}(\tau)/\epsilon_0\right)^{-1/4}$, 
$\lambda_1(\tau)=\epsilon_0\lambda_1^0\left(T_{\tau\tau}(\tau)/\epsilon_0\right)^{1/2}$~\cite{BRSSS}. Here, the stress tensor is that on the local rest frame (LRF) of the Bjorken flow and it is diagonal. Other components of the stress tensor, $T_{yy}$ and $T_{x_{\perp}x_{\perp}}$ where $y$ is the rapidity and $\vec{x}_{\perp}$ denote the perpendicular directions to the collisional axis, are expressed  in terms of $T_{\tau\tau}$ by virtue of the EOS and the hydrodynamic equation.

We propose the following parametrization of the five-dimensional (5d) dual metric on the Eddington-Finkelstein coordinates:
\begin{align}
ds^{2}=	&-r^{2}a\,d\tau^{2}
		+2\,d\tau dr
		+e^{2b-2c}\left(1+r\tau\right)^{2}dy^{2}\nonumber\\
	&\quad
		+r^{2}e^{c}\,d\vec{x}_{\perp}^{2}.
\label{Fin-assum}
\end{align}
Here, $r$ is the fifth dimension with $r\to\infty$ corresponding to the spatial boundary. 
The boundary conditions at $r\to\infty$ are set so that the 4d part of the metric becomes $r^{2}$ times the 4d Minkowski metric on the LRF of the Bjorken flow:
\begin{align}
ds^{2}|_{r\to\infty}
=	r^{2}\left(
-d\tau^{2}+\tau^{2}dy^{2}+d\vec{x}_{\perp}^{2}\right)
		+2\,d\tau dr,
		\label{boundary_condition}
\end{align}
namely $a\to 1$ and $b,c\to 0$. We assume that the bulk metric depends only on $\tau$ and $r$ to ensure the boost invariance ($y$-independence)~\cite{Bjorken}. We have also assumed that the 4d part of the bulk metric is diagonal so that the diagonal 4d stress tensor is reproduced.  Physically, these conditions provide information to the gravity theory that the plasma undergoes the Bjorken flow. We note that there remains a gauge degree of freedom $r\to r+f(\tau)$ in the metric~\eqref{Fin-assum}.

$a,b,c$ (hence the 4d stress tensor) are determined from
the vacuum Einstein equation with the negative cosmological constant ($\Lambda=-6$ in our unit), which is equivalent to (the bosonic part of) the equations of motion of 10d type IIB supergravity under the assumption of constant dilaton and constant Ramond-Ramond flux~\footnote{Constant dilaton and constant Ramond-Ramond flux together with the solution presented in this paper solve the original 10d equations of motion.}. 
Eq.~(\ref{T_four_D}) suggests that the Einstein equation may also be solved order by order in the $\tau^{-2/3}$ expansion. 
A natural trial suggested by Ref.~\cite{Janik-Pes} is that we introduce a new spatial variable $u\equiv r\tau^{1/3}$ and perform a $\tau^{-2/3}$ expansion regarding that $u$ is independent of $\tau$: 
$
a(u,\tau)
	=	a_0(u)+a_1(u)\tau^{-2/3}+a_2(u)\tau^{-4/3}+\cdots .
$
(This is similar for $b$ and $c$.)
We shall find that this expansion (which we call late-time expansion in this Letter) works consistently.

With the boundary conditions (\ref{boundary_condition}), we find the following solutions to the Einstein equation:
\begin{eqnarray}
a_{0}(u)&=&1-w^{4}u^{-4},
\quad b_{0}(u)=c_{0}(u)=0,
\nonumber \\
a_{1}(u)&=&-\frac{2}{3}\frac{(1+\xi_{1})u^{4}+\xi_{1} w^{4}-3\zeta_1 u w^{4}}{u^{5}},
\nonumber \\
b_{1}(u)&=&-(\xi_{1}+1)/u,
\nonumber \\
c_{1}(u)
&=&\frac{1}{3w}
\left[
\arctan{u\over w}-\frac{\pi}{2}+\frac{1}{2}\log\left(\frac{u-w}{u+w}\right)
\right]
\nonumber \\
&&\quad
-\frac{\zeta_1}{2}\log\left(1-{w^4\over u^4}\right)
-\frac{2\xi_{1}}{3u},
\label{first-solution-general}
\end{eqnarray}
where $\xi_1$, $w$, $\zeta_1$ are the integration constants which are not determined by the boundary conditions.
We have also obtained the second-order solution, which is too lengthy to be shown here and will be presented in Ref.~\cite{ours}. The second-order solution contains two more integration constants $\xi_2$ and $\zeta_2$.
Actually, we can show that $\xi_{1}$ and $\xi_2$ are gauge degrees of freedom which can be absorbed by the following coordinate transformation:
\begin{align}
u\to u-3^{-1}\xi_1\tau^{-2/3}-3^{-1}\xi_2\tau^{-4/3}+O(\tau^{-2}). 
\label{coord_tr}
\end{align}
There is also a gauge degree of freedom of constant shift $u\to u+\xi_0$, but
we have already chosen a gauge $\xi_0=0$ in (\ref{first-solution-general}).


Through the standard AdS/CFT dictionary~\cite{BK}, we find that the integration constants $w$, $\zeta_1$ and $\zeta_2$ are related to the energy density of the 4d fluid $T_{\tau\tau}$ by
\begin{align}
	T_{\tau\tau}
		&=	{3w^4\over16\pi G_5}\left(\tau^{-4/3}-2\zeta_1\tau^{-2}+\zeta_2\tau^{-8/3}+\cdots\right).
			\label{T_five_D}
  \end{align}
Other non-zero components $T_{yy}$ and $T_{x_{\perp}x_{\perp}}$ are expressed in terms of $T_{\tau\tau}$ by virtue of the Einstein equation at the vicinity of the boundary~\cite{ours} in a consistent way with hydrodynamics. Namely, the hydrodynamic equation and EOS are encoded in the Einstein equation and the dual geometry~\footnote{This fact has been observed earlier in Refs.~\cite{Friess:2006fk,BHMR}.}. 
%
Comparing Eqs.~\eqref{T_five_D} and \eqref{T_four_D}, we can read off the physical meaning of the integration constants as $\zeta_1=\eta_0$ and $\zeta_2=\epsilon_0^{(2)}$.

{\it Regularity of dual geometry to all orders}.---The Kretschmann scalar computed from our metric to the first order is
$	(R_{\mu\nu\rho\sigma})^2
		=	40+72w^8/u^8+O(\tau^{-2/3})$.
Obviously there is a singularity at the origin $u=0$.
We can also compute a component of the Riemann tensor projected onto an orthonormal basis
\begin{align}
	R^{\underline{y}}{}_{\underline{4}\underline{y}\underline{4}}
		 =
	R^y{}_{\mu y\nu}e^\mu_{\underline{4}} e^\nu_{\underline{4}}
		&=	{w^4\over3u^2(u^4-w^4)^2}\left(\zeta_1-{4u^3\over3(3u^4+w^4)}\right)\nonumber\\
		&\quad
			+O(\tau^{-2/3}),
			\label{Riemann_tetrad}
  \end{align}
where $e^\mu_{\underline{4}}=-2^{-1/2}(1,0,0,0,1+2^{-1}r^2a)$ is a fifth-dimensional component of the f\"unf\-bein.
The expression~\eqref{Riemann_tetrad},
which is a coordinate scalar (and a local Lorentz tensor), is singular at $u=w$ except when
\begin{align}
\zeta_1={1/3w}.
	\label{first_reg_cond}
  \end{align}
Hence, the spacetime regularity at $u=w$ requires~\eqref{first_reg_cond}.

Similarly, we can show that the regularity of $R^{\underline{y}}{}_{\underline{4}\underline{y}\underline{4}}$ at $u=w$ is achieved at the second order if~\cite{ours}
\begin{align}
\zeta_2={(1+2\log2)/18w^2}.
	\label{second_reg_cond}
  \end{align}
Recalling the correspondences $\zeta_1=\eta_0$ and $\zeta_2=\epsilon_0^{(2)}$, one can see that the regularity of the spacetime has completely fixed the shear viscosity and the combination of the second-order transport coefficients~\footnote{The same results are obtained by requiring the regularity of the Kretschmann scalar at one order higher for each, namely at the second order for $\eta_0$~\cite{Janik-eta} and at the third order for $\epsilon_0^{(2)}$~\cite{HJ}.}.
Let us assume that the static relation between the energy density $T_{\tau\tau}$, the entropy density $s$ and the temperature $T$~\cite{Gubser:1996de} is still valid due to the local thermal equilibrium which is realized at a sufficiently large proper time: $s(\tau)={\pi^2N_c^2\over2}T^3(\tau)$.
Then we obtain, by using $N_c^2/4\pi^2=1/8\pi G_5$, that ${\eta(\tau)/s(\tau)}={3\eta_0w/4\pi}$ which results in the celebrated formula~\cite{viscosity}
$	{\eta(\tau)/ s(\tau)}
		=	{1/4\pi}$,
as a consequence of the regularity condition~\eqref{first_reg_cond}.
Similarly we get from the second-order results that
$	\lambda_1(\tau)-\eta(\tau)\tau_\Pi(\tau)
		=	-{1-\log2\over9}N_c^2\left(T_{\tau\tau}(\tau)/s\right)^2$,
which is consistent with the results in \cite{HJ,BRSSS,BHMR,NO}. 

So far, the conditions~\eqref{first_reg_cond} and \eqref{second_reg_cond} are the necessary ones for the spacetime regularity at $u=w$.
Actually, we find that these are the sufficient conditions, too. This can be shown by confirming the regularity of the metric and its inverse as well as their arbitrary-order derivatives with respect to $r$ and $\tau$. (The apparent divergence of the metric at $r\to\infty$ is harmless as it is the case for the pure AdS.)
Once these are shown to be regular, any coordinate invariants made of Riemann tensors and their covariant derivatives are also regular. The foregoing regularity conditions are equivalent to the regularity of $a, b, c$ and their arbitrary-order derivatives in our parametrization (\ref{Fin-assum}) except at the origin.

We can generalize the above observation to all orders by means of induction~\cite{ours}.
The outline is the following. Let us use the terminology ``regular'' for the regularity except at the origin. We begin with the assumption that $a_{k}$, $b_{k}$, $c_{k}$ and their arbitrary-order derivatives are regular for $k<n$. We also assume that the $1/u$ expansions of $a_{k}$, $b_{k}$, $c_{k}$ around the boundary start at the order of $1/u$ or less singular order. Then, a component of the $n$th order ($\propto\tau^{-2n/3}$) Einstein equation,
\begin{align}
	(u^2b_n')'
		&=	\text{terms including $a_k$, $b_k$, $c_k$ ($k<n$) and}\nonumber\\
		&\qquad
			\text{their derivatives with regular coefficients},
		\label{b_regularity}
  \end{align}
tells us that the left-hand side is regular, where $'$ denotes $\partial/\partial u$. We can generalize this statement to the regularity of $b_{n}$ and its arbitrary-order derivatives for $0<u\leq\infty$ by integrating and/or differentiating the equation~\eqref{b_regularity}. 
We can also prove the regularity of $a_{n}$ and its arbitrary-order derivatives in a similar way, by using another component of the Einstein equation:
\begin{align}
	(u^4a_n)'
		&=	\text{terms including $a_k$, $b_k$, $c_k$ ($k<n$), $b_n'$, and}\nonumber\\
		&\qquad
			\text{their derivatives with regular coefficients}.
  \end{align}

The proof for the regularity of $c_{n}$ is more complicated since we encounter a potential singularity:
\begin{align}
	c_n'	&=	-{2\over3}b_n'+{1\over u^4-w^4}\left(\text{regular terms}\right).
		\label{c_regularity}
  \end{align}
It is necessary to prove that one can take the integration constants so that the terms denoted as ``regular terms'' in Eq.~\eqref{c_regularity} vanishes at $u=w$.
We find that the integration constant~$\zeta_n$ in $a_{n}$ appears linearly in the regular terms of Eq.~\eqref{c_regularity} hence it is always possible to remove the singularity by an appropriate choice of $\zeta_n$. We also find that $\zeta_n$ corresponds to the $n$th order contribution to the stress tensor~$\epsilon_0^{(n)}$. We have explicitly seen this to the second order: the transport coefficients $\eta_{0}$ and $\epsilon_0^{(2)}$ are determined by requesting the regularity of $c_{1}$ and $c_{2}$, respectively. 
Since we have already shown that our starting assumption is valid to the second order, the regularity (under the appropriate choice of the transport coefficients) for all orders is proven by induction.

So far we have proven that the spacetime can be made regular by an appropriate choice of the integration constants $\zeta_n$ that corresponds to the transport coefficient $\epsilon_0^{(n)}$.
Now we shall prove that if we do not choose the constants $\zeta_n$ as above, 
we necessarily encounter a singularity at $u=w$.
This is shown by computing
$	R^{\underline{y}}{}_{\underline{4}\underline{y}\underline{4}}
		=	{1\over72\tau^{-2/3}u^4(\tau^{-2/3}+u)}\left({c_n'\over u}+{c_n''\over2}+\text{regular terms}\right)$.
Any choice which makes Eq.~\eqref{c_regularity} singular at $u=w$ necessarily renders $R^{\underline{y}}{}_{\underline{4}\underline{y}\underline{4}}$ singular.

We have proven that a unique appropriate choice of a set of transport coefficients makes the spacetime regular (except at the origin).
Therefore, the unremovable singularity pointed out in Ref.~\cite{BBHJ} is absent in our setup.

{\it Apparent horizons}.---Since we still have a physical singularity at the origin ($u=0$),
we shall show the presence of an event horizon which covers it.
%
We show the presence of an apparent horizon whose existence necessarily leads to the existence of an event horizon outside~\cite{Hawking-Ellis}. 
Let us expand the location of the apparent horizon as $u_{H}(\tau)=u_0+u_1\tau^{-2/3}+u_2\tau^{-4/3}+O(\tau^{-2})$ and determine the coefficients order by order by solving the equation
$
0=\Theta(u_{H})
=\Theta_{0}+\Theta_{1}\tau^{-2/3}
+\Theta_{2}\tau^{-4/3}+O(\tau^{-2}),
$
where $\Theta$ is the normalized product of expansions in the double null formalism~\cite{double-null}. We obtain
%
\begin{align}
\Theta_{0}	&=	-(9/2)(1-u_0^{-4}w^{4}),\\
\Theta_{1}	&=	-(3/w)(1-2\xi_{1}+6u_{1}),\\
\Theta_{2}	&=	-\frac{3 \pi+4 \log 2-8-24 w \left(\xi _2-3 u_2\right)}{4 w^2},
\end{align}
where the conditions for the regularity~\eqref{first_reg_cond} and \eqref{second_reg_cond} have been imposed.
The apparent horizon is located at
$u_0=w$, $u_1=-\frac{1}{6}+\frac{\xi_{1}}{3}$ and 
$u_2=\frac{8-3\pi-4\log2}{72w}+\frac{\xi_{2}}{3}$.
We see the presence of the trapped region $\Theta>0$ inside the apparent horizon.
The contribution of $\xi_{1}$ and $\xi_2$ can be absorbed by the coordinate transformation~(\ref{coord_tr}).

The area element of the apparent horizon is given by $A_H dy d\vec{x}_{\perp}^{2}$, where
$	A_H	=w^3-{w^2\over2}\tau^{-2/3}
+{w(2+\pi+6\log2)\over24}\tau^{-4/3}+O(\tau^{-2})$.
The area element of the event horizon should be the entropy density per unit rapidity $\tau s$, that means, $\tau s\geq A_H/4G_5$.
(At the late-time limit, the apparent horizon must coincide with the event horizon.)
The entropy density per unit rapidity is given by $\tau s\propto \tau T_{\tau\tau}^{3/4}\propto 1-{3\eta_0\over2}\tau^{-2/3}+\cdots$ from Eq.~\eqref{T_four_D} and the dimensional analysis.
The entropy production rate coincides with the expansion rate of the area of the apparent horizon at the leading order:
${A_H^{-1}\partial_\tau A_H}={(\tau s)^{-1}\partial_\tau (\tau s)}
=\eta_{0}\tau^{-5/3}$.
In other words, the ratio of the first and the second terms in the $\tau^{-2/3}$ expansion agrees between $A_{H}$ and $\tau s$.

The area element of the apparent horizon has been found to increase with $\tau$.
Further, we can show that the difference between the Weyl tensors exhibits a time-dependent anisotropy:
$	C^{x^{1}x^{2}}{}_{x^{1}x^{2}}-C^{x^{1}y}{}_{x^{1}y}
		=	{3\eta_0w^4  u^{-4}}\tau^{-2/3}+O(\tau^{-4/3})$.
This indicates that the geometry is not locally static under the presence of the dissipation due to the viscosity.

{\it Summary and discussion}.---We have proposed a gravity dual of the Bjorken flow of the $\mathcal{N}=4$ SYM fluid.
The existence of event horizon has been shown for the first time in the geometry dual to the Bjorken flow.
We have also proven that the spacetime can be made regular to all orders of the late-time expansion by an appropriate choice of the integration constants corresponding to the viscosity, the relaxation time, and other transport coefficients.
Furthermore, this choice is not only sufficient for the regularity but also necessary, namely, the requirement of the regularity completely fixes these coefficients.
The area element of the apparent horizon has also been computed.
The increasing rate of the area element coincides with that of the entropy density per unit rapidity at the leading order.
Our spacetime exhibits a time-dependent anisotropy in the Weyl tensors, indicating that the geometry is not locally static.

It is interesting to consider what classifies the properties determined around the boundary and those determined around the horizon. The EOS and the hydrodynamic equation have been obtained at the vicinity of the boundary while the transport coefficients have been obtained from the regularity around the horizon. The traceless property (EOS) and the conservation of the stress tensor (hydrodynamic equation) hold whether or not the (local) thermal equilibrium is achieved in the YM-theory side. However, the concept of the transport coefficients makes sense only when the notion of fluid is valid. This tempts us to relate the notion of (local) thermal equilibrium with the regularity (or the presence) of the horizon. 
It is also interesting to see how the method to determine the transport coefficients from the regularity is related to the Kubo's linear response theory (Kubo formula) and other holographic computations. 
We hope that these points will be clarified in the future.

We have discussed the time evolution of the entropy density from the viewpoint of the dual geometry. It would also be interesting to study a thermodynamic formulation of dynamical black holes in asymptotically $\text{AdS}_{5}$ geometries. 

The present paper provides a consistent setup for the holographic dual of the Bjorken flow of the ${\cal N}=4$ SYM plasma. Our model also provides a concrete well-defined example of time-dependent AdS/CFT. We hope that the present work sheds some light on the dynamical nature of the time-dependent plasma.

{\it Note added}:
When the present work was at the final stage, we received a paper~\cite{Heller} which partly covers a related subject.
The first-order solution presented in Ref.~\cite{Heller} corresponds to the gauge choice of $\xi_1=0$ in our first-order solution. Our proposal in the present paper has been invented independently. However, we were motivated by Ref.~\cite{Heller} to examine the gauge degree of freedom and the regularity of the higher-order geometry.

{\it Acknowledgement}:
We would like to thank Masayuki Asakawa, Akihiro Ishibashi and Makoto Natsuume for useful discussions. 
This work was initiated during the YITP-W-06-11 on ``String Theory and Quantum Field Theory,'' and discussions during the YITP international symposium ``Fundamental Problems in Hot and/or Dense QCD'' were useful. The authors also thank APCTP where useful discussions have been made during the focus program ``New Frontiers in black hole physics.''
This work is partially supported by MEXT, Japan, 
JSPS (S.K.), WPI Initiative, Nos.~17740134, 19GS0219, 19340054 (S.M.), 19740171, 20244028, and 20025004 (K.O.) and by KOSEF, Korea, Grant Nos.~R01-2004-000-10520-0 and R11-2005-021 (S.N.).

\end{document}